\documentclass[12pt]{article}
\usepackage{pic04}
\usepackage{hyperref}
\usepackage{url}
\usepackage{graphicx}

\def\mvC     {{\mbox{$\mathrm{MeV/c^2}$}}}
\def\mvc     {{\mbox{$\mvC \ $}}}

\def\dsj     {{\mbox{$\mathrm{D^{+}_{sJ}}$}}}

\def\dse     {{\mbox{$\dsj(2632) \rightarrow \rm{D_{s}^{+}} \eta \ $}}}
\def\dzk     {{\mbox{$\rm{D^{0} K^+} $}}}
\def\dseta   {{\mbox{$\rm{D_{s}^{+} \eta} \ $}}}

\def\Nevt    {101 }                             
\def\Back    {54.9 }                            
\def\Mass    {2635.4 }                          
\def\sMass   {3.3 }                             

\def\Nevtb   {21 }                              
\def\Backb   {6.9 }                             
\def\Massb   {2631.5 }                          
\def\sMassb  {2.0 }                             

\def\MassA   {2632.5 }                          
\def\sMassA  {1.7 }                             

\def\Br      {0.14}                             
\def\dBr     {0.06}                             

\begin{document}

\title{\bf First Observation of a New Narrow \dsj\  Meson at 2632 \mvc}

\author{Peter S. Cooper\\
        {\em Fermi National Accelerator Laboratory}
        \\(For the Selex Collaboration)
        \thanks{http://www-selex.fnal.gov/}}
\maketitle

%
%
%
%
%
%
\vspace{4.5cm}
%

\baselineskip=14.5pt
\begin{abstract}
We report the first observation of a charm-strange meson 
$\dsj(2632)$ at a mass of $\MassA \pm \sMassA\ \mvc$ in data from SELEX, 
the charm hadro-production experiment E781 at Fermilab.  This state is seen in 
two decay modes, \dseta and \dzk.  In the $\rm{D_{s}^{+} \eta}$ decay mode 
we observe a peak with \Nevt events over a combinatoric background of \Back 
events at a mass of \Mass $\pm$ \sMass \mvC.  There is a corresponding peak of 
\Nevtb events over a background of \Backb at \Massb $\pm$ \sMassb \mvc in the 
decay mode \dzk.  The relative branching ratio $\Gamma(\dzk)/\Gamma(\dseta)$ 
is \Br \ $\pm$ \dBr.  The full vesion of this paper has been accepted for 
publication in PRL~\cite{Evdokimov:2004iy}.  Here I have reproduced only the 
mass difference signal plots for the \dseta and \dzk decay modes.
\end{abstract}
\newpage

\begin{figure}[ht]
\includegraphics[width=3.1in]{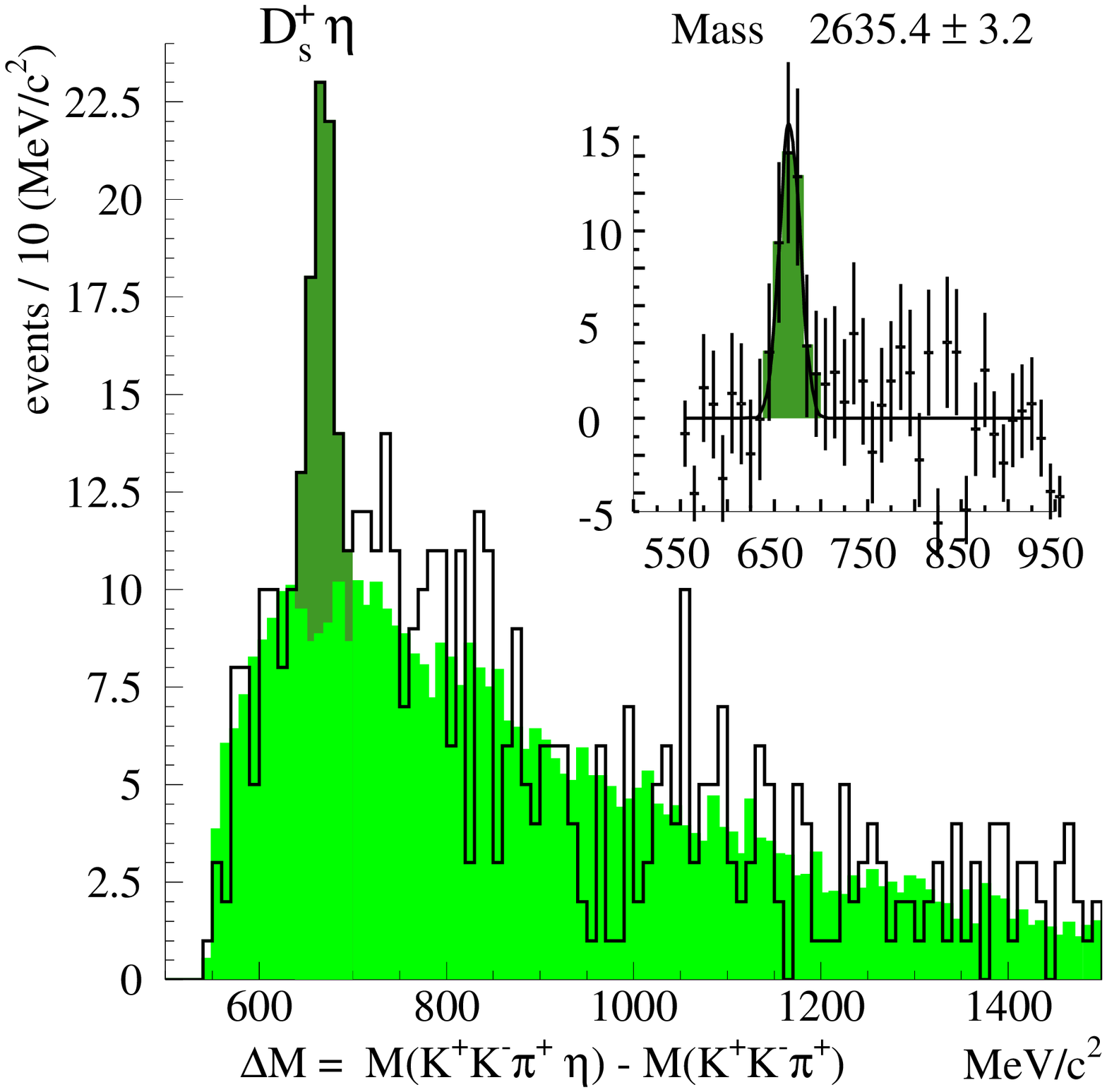}
\label{dseta}
\end{figure}

\begin{figure}[ht]
\includegraphics[width=3.1in]{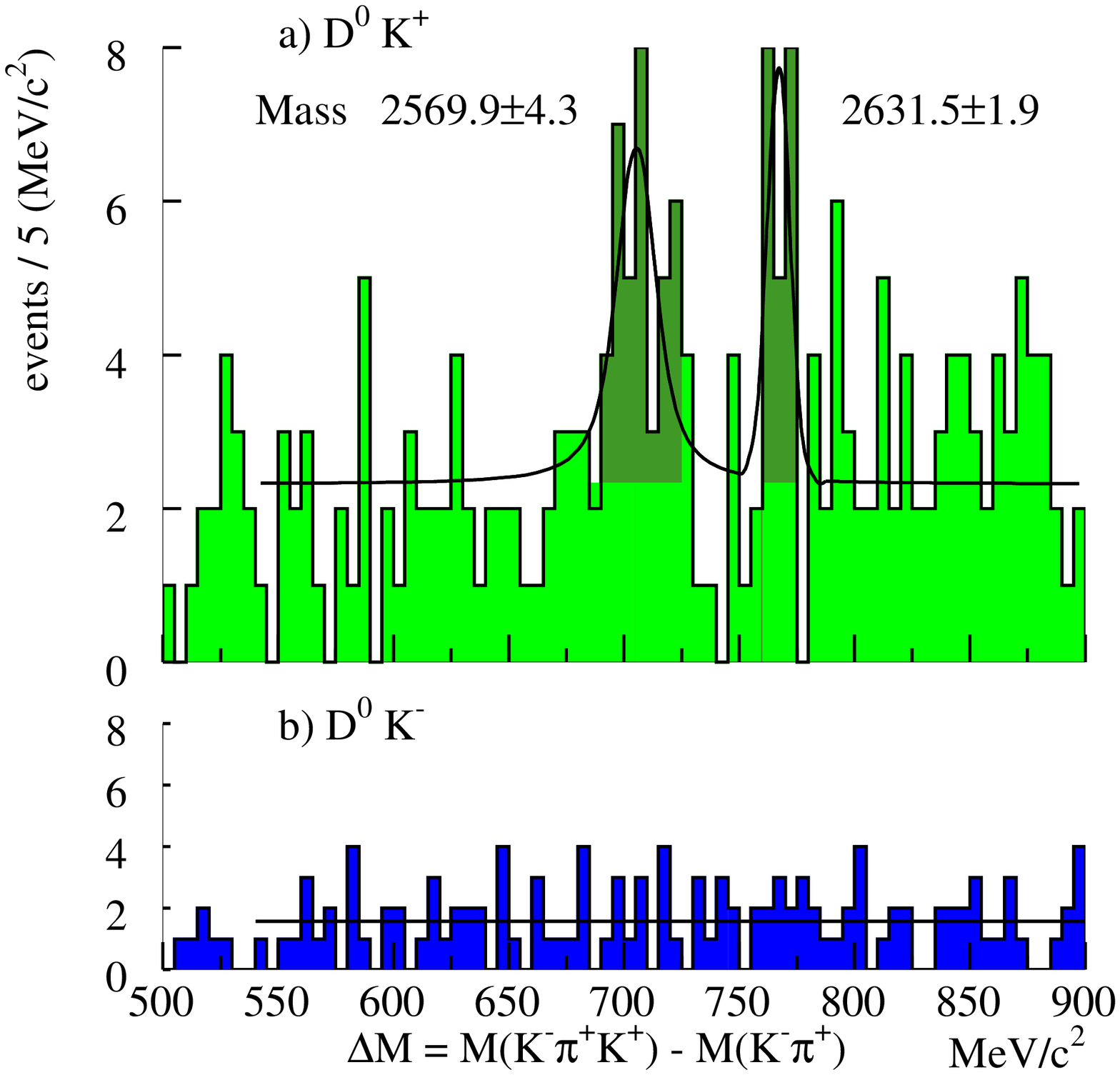}
\label{d0k}
\end{figure}

\end{document}